\documentclass[prl,onecolumn,showpacs,epsfig,longeq]{revtex4}
\usepackage{graphicx}
\usepackage{dcolumn}
\usepackage{amssymb}
\usepackage{bm}
\usepackage{epsfig}
\usepackage{psfrag}

\def\3dots{\:\raisebox{-0.5ex}{$\stackrel{\textstyle.}{:}$}\:}
\def\beq{\begin{equation}}
\def\eeq{\end{equation}}
\def\bea{\begin{eqnarray}}
\def\eea{\end{eqnarray}}

\begin{document}

\title{
Evolutionary Rotation in Switching Incentive Zero-Sum Games
}
\author{ Zhijian Wang$^{a}$\footnote{Corresponding author: wangzj@zju.edu.cn. The two author contribution equal.}}
\author{Bin Xu$^{b,a}$ }
\affiliation{$^{a}$Experimental Social Science Laboratory, Zhejiang University, Hangzhou 310058, China}
\affiliation{$^{b}$College of Public Administration, Zhejiang Gongshang University, Hangzhou 310000, China}

\date{\today}
\begin{abstract}
In a competitive game situation where the only Nash equilibrium is in mixed strategies, persistent cyclic social rotation can be expected. The rotation is pointed out firstly by Shapley (1964) as never-ending cycle and in discrete condition called as Shapley-ploygon. To insight into the rotation is long time expectation.
Data comes from an 13 groups, 825 rounds and 7 games experiment, which are of random matching two-persons zero-sum extended 2$\times$2 games (Binmore, Swierzbinski and Proulx, 2001~\cite{Binmore2001}). Whenever the payoff matrix is switched by the experimenters,
 we find, the human subjects social rotations' directions and strengths is changed simultaneously. The directions of rotation can be captured by evolutionary dynamics; meanwhile, an explanation for observed rotation strengths is also given with eigenvalues calculated from  the Jacobian of normalized replicator dynamics. In addition, equality-of-populations rank test shows that relative response coefficients within groups differ and persist cross the parameters switching.
In the old data, we demonstrate a new view --- rotation --- to insight the Shapley-cyclic motion strength   quantitatively.
\end{abstract}
\maketitle


\section{1. Introduction}
%

In theoretical economic history,  Shapley's the never-ending cycle in game (Shapley, 1964), as the Edgeworth cycle in price (Edgeworth, 1925), is a deep insight into stochastic processes of social evolution~\cite{benaim2009learning}.  Shapley never-ending cycle (now called as Shapley-polygon) state that, in a competitive situation where the only Nash equilibrium is in mixed strategies, persistent cycles can be expected~\cite{benaim2009learning}. The unstable equilibrium has been strongly linking to evolutionary game theory (EGT) and learning processes~\cite{Hofbauer1995,benaim2009learning,Cason2010} in the past decades.

Experimental economics can be a method for theory testing~\cite{Falk2009,Samuelson2002}. To the best of our knowledge, the Shapley-polygon has rare been well experimental tested. Only closing experimental literature comes from Cason, Friedman and Hopkins~\cite{Cason2010}. The strength of rotation of Shapley-polygon has never been reported. In this letter, in old experimental data of zero-sums games~\cite{Binmore2001}, we report that, not only the direction, but also the strength of social rotation can be distinguished, and even more, quantitatively, the results support the predictions from Shapley-polygon and EGT.

\begin{figure}
---------------------------------------------------------------------------------------------------------------------------------------------------------\\
\textbf{(a)}\\
---------------------------------------------------------------------------------------------------------------------------------------------------------\\
\centerline{\includegraphics[width=0.70\textwidth]{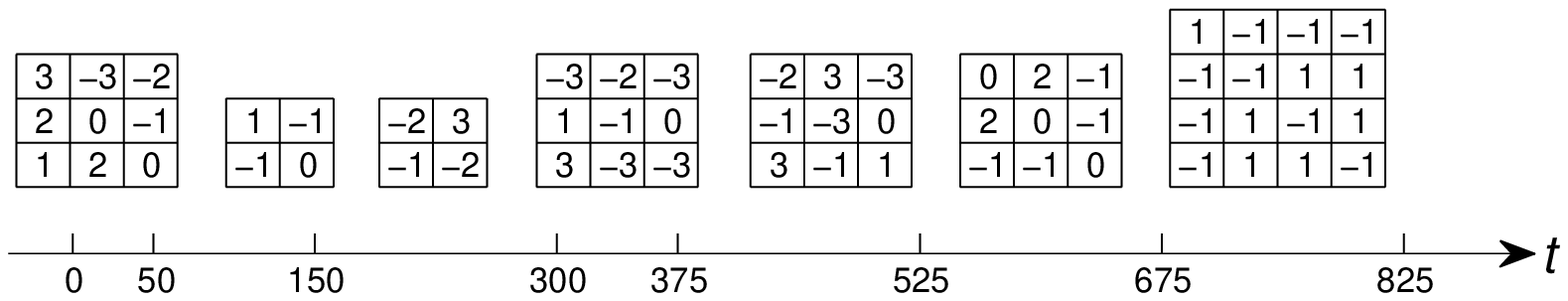}}
---------------------------------------------------------------------------------------------------------------------------------------------------------\\
\textbf{(b)}\\
---------------------------------------------------------------------------------------------------------------------------------------------------------\\
\centerline{\includegraphics[width=0.70\textwidth]{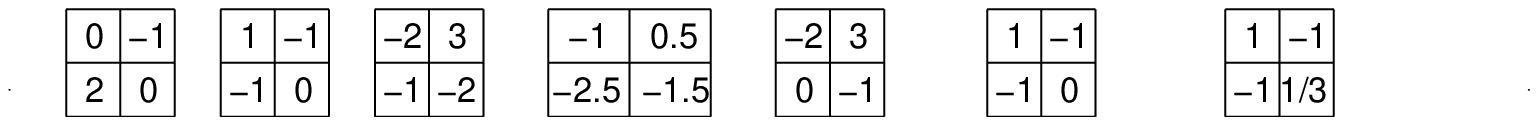}}
---------------------------------------------------------------------------------------------------------------------------------------------------------\\
\textbf{(c)}\\
---------------------------------------------------------------------------------------------------------------------------------------------------------\\
\centerline{{\includegraphics[width=1.6cm]{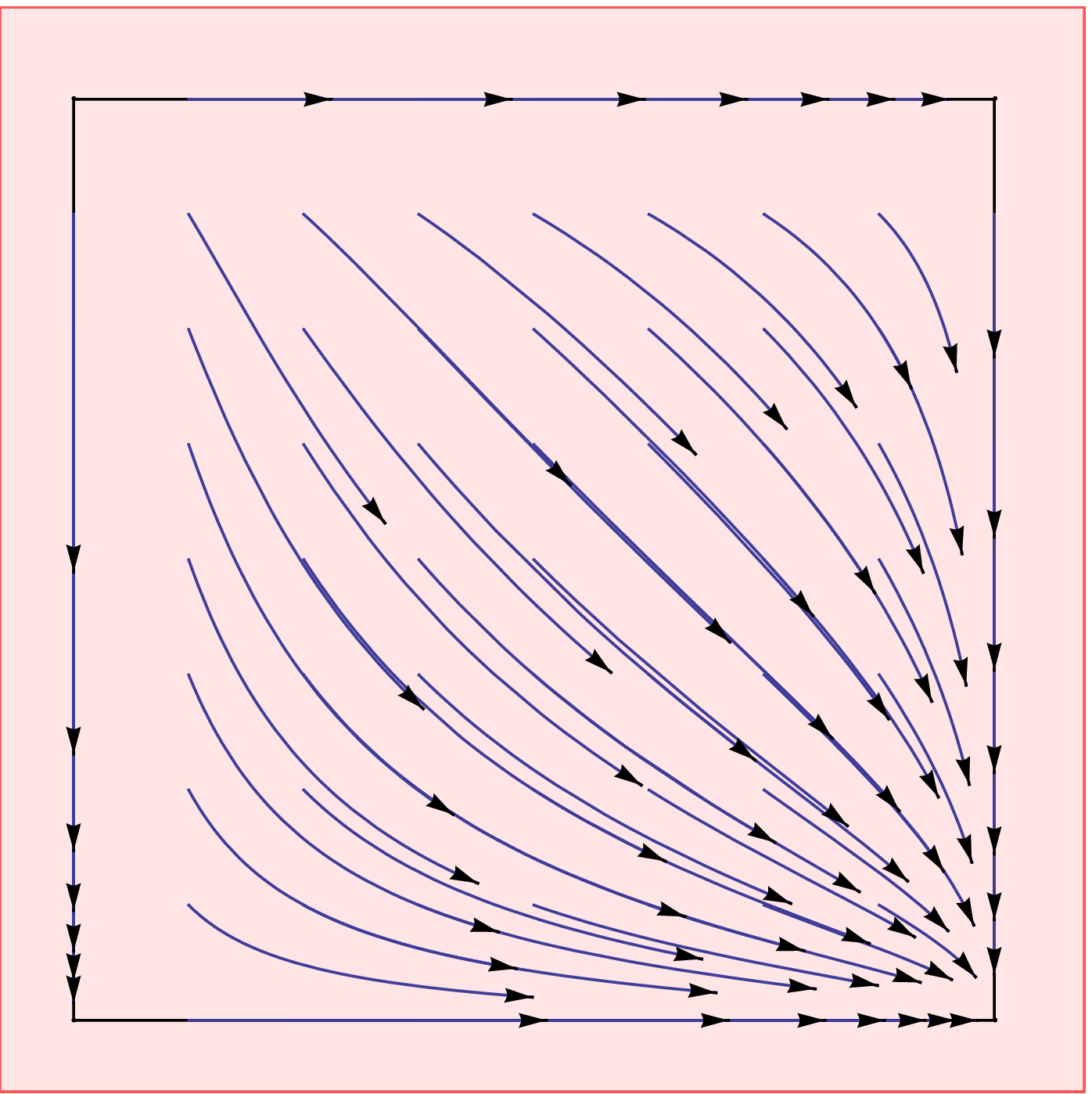}}
{\includegraphics[width=1.6cm]{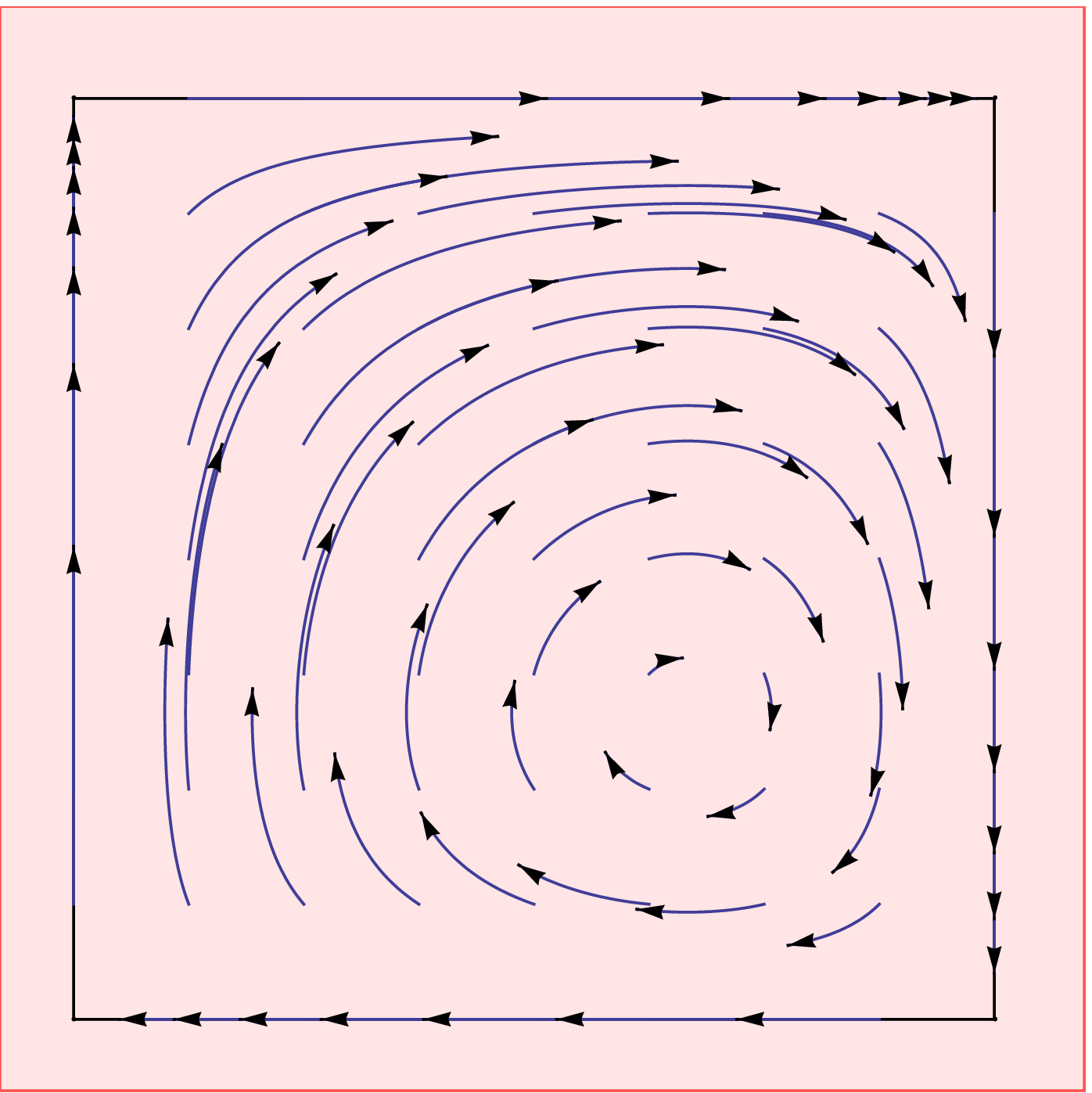}}
{\includegraphics[width=1.6cm]{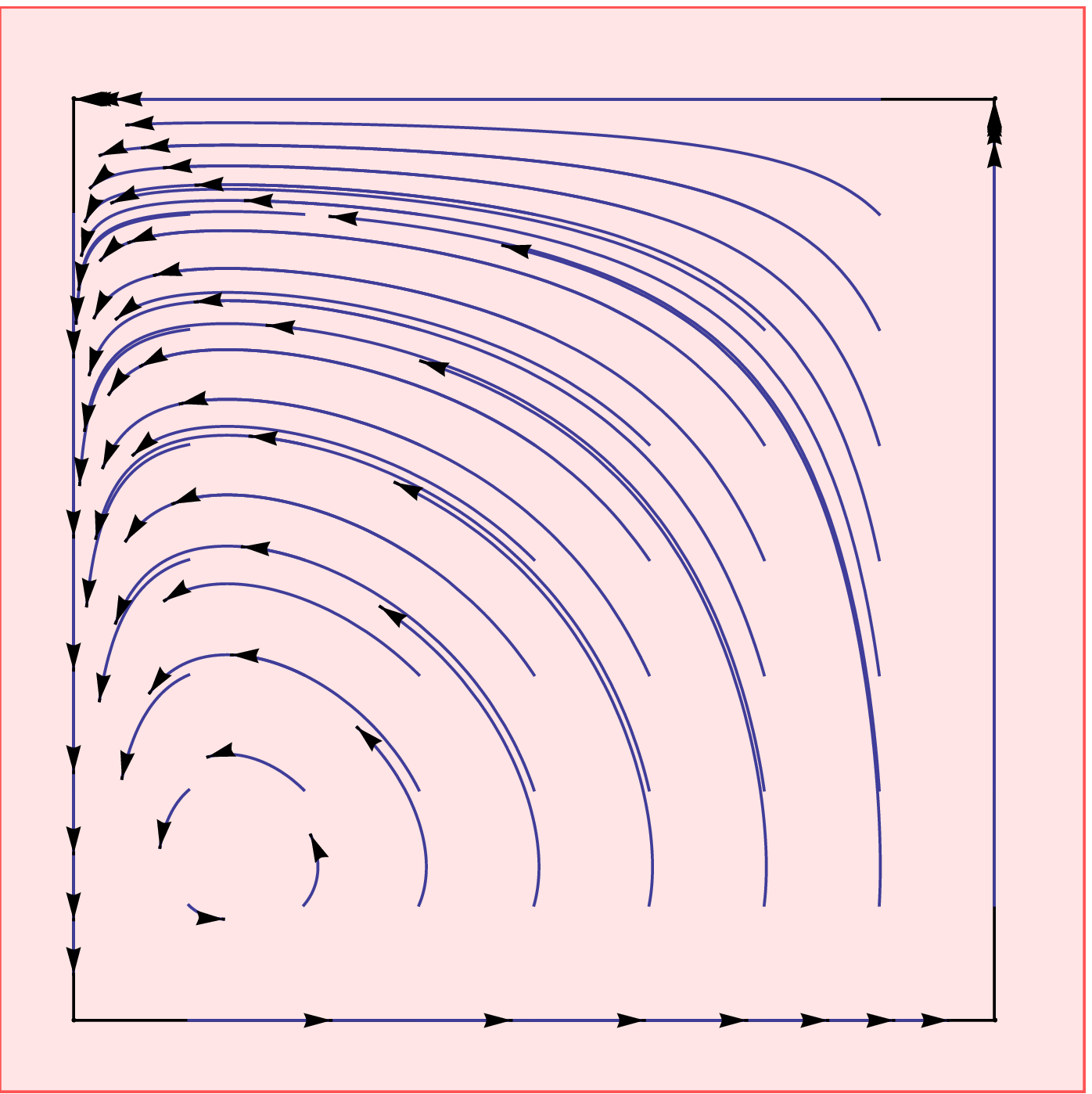}}
{\includegraphics[width=1.6cm]{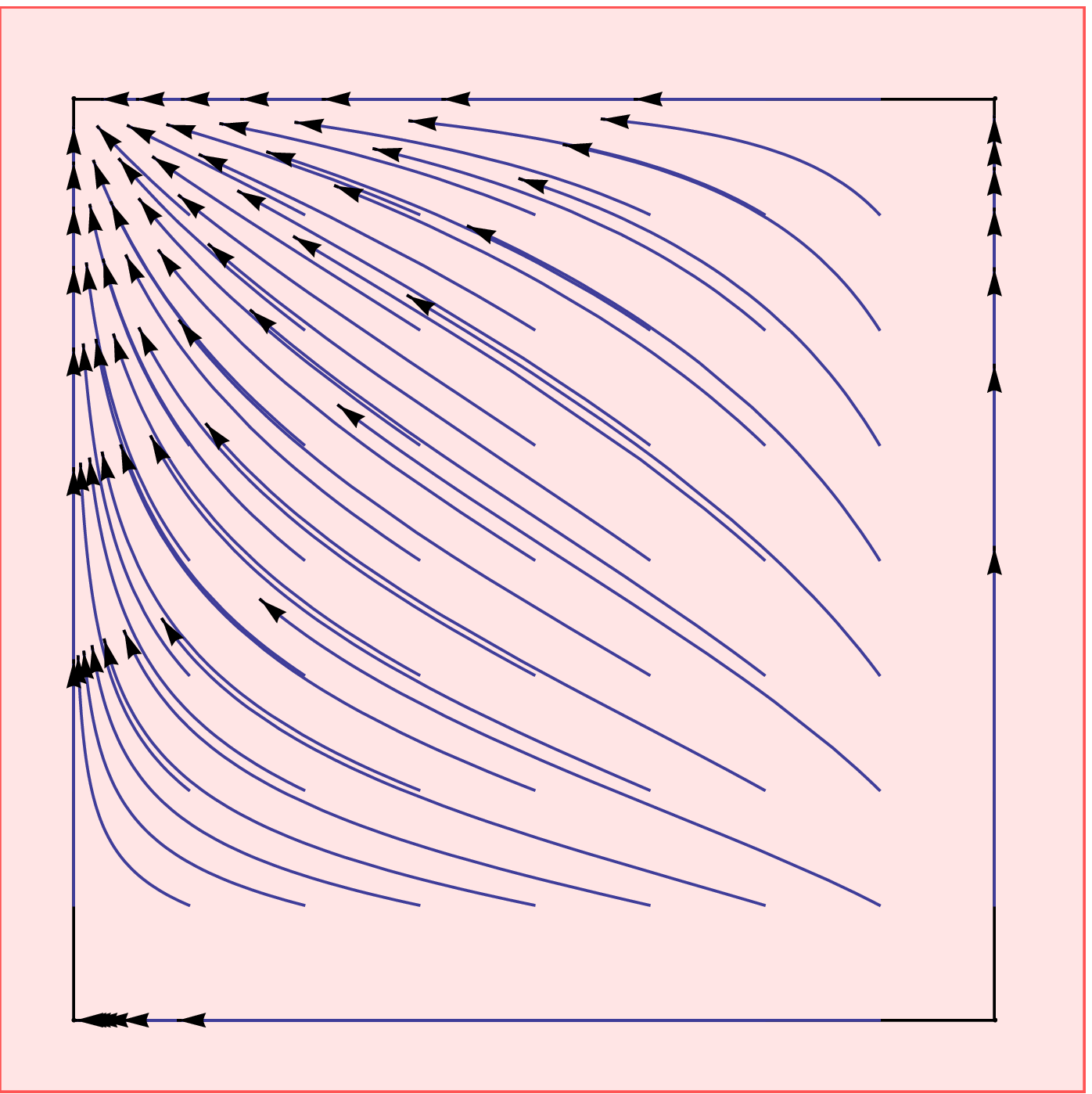}}
{\includegraphics[width=1.6cm]{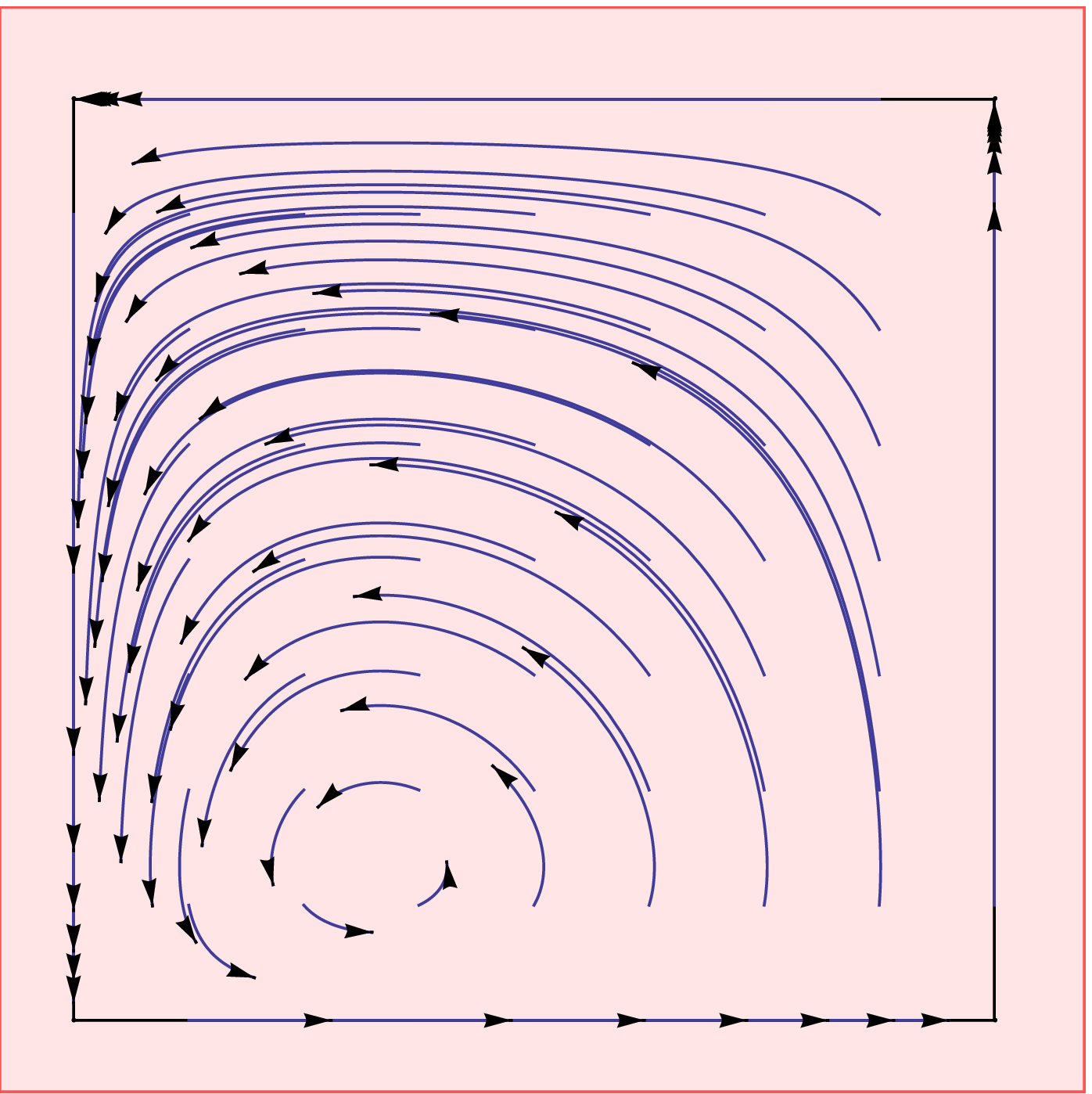}}
{\includegraphics[width=1.6cm]{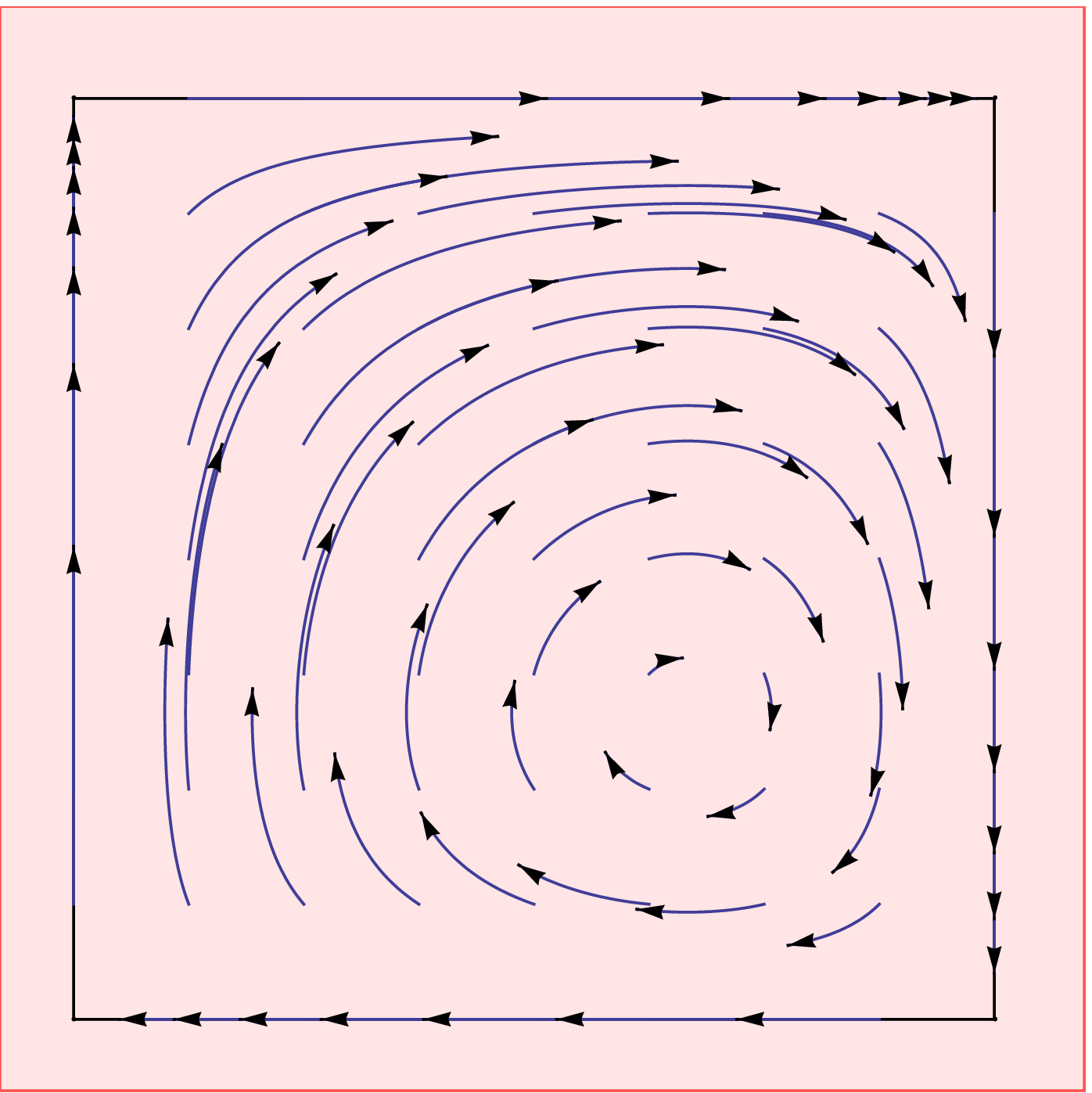}}
{\includegraphics[width=1.6cm]{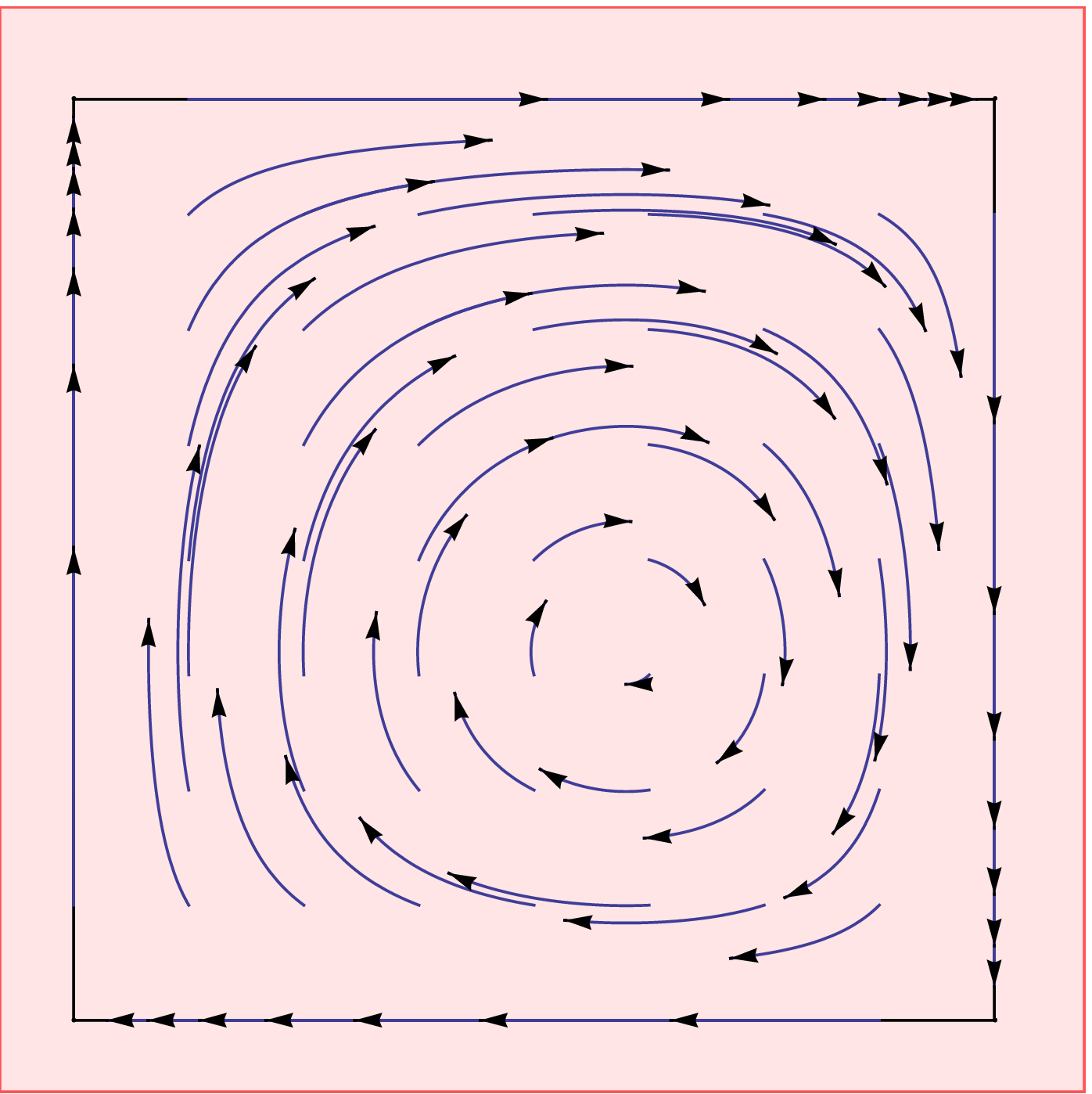}}}
~~~~~~~~~~~~~~~~~~~~~~~~~~~~~\textbf{0}~~~~~~~~~~~~~~\textbf{-}~~~~~~~~~~~~~~\textbf{+}~~~~~~~~~~~~~~\textbf{0}~~~~~~~~~~~~~\textbf{+}~~~~~~~~~~~~~\textbf{-}~~~~~~~~~~~~~~~\textbf{-}~~~~~~~~~~~~~~~~~~~~~~~~~~~~~\\
---------------------------------------------------------------------------------------------------------------------------------------------------------\\
\textbf{(d)}\\
---------------------------------------------------------------------------------------------------------------------------------------------------------\\
\centerline{\includegraphics[width=0.75\textwidth]{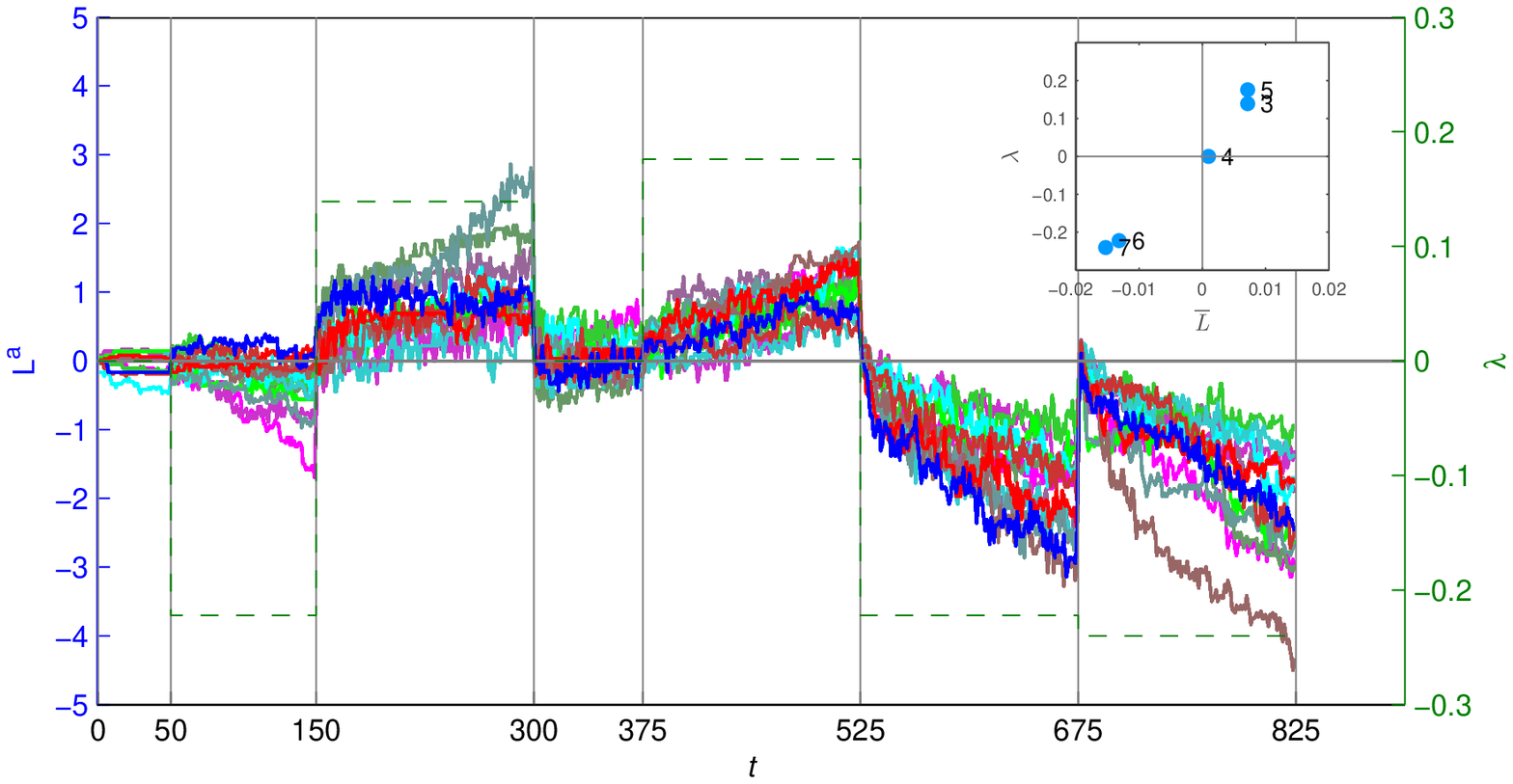}}
\caption{Illustration of the Experiments.  Each of the 13 groups ($\beta=1,2,...,13$) play 7 games ($\alpha=1,2,...,7$) and 825 periods total. (a) From left to right, the matrix are the payoff matrix of the 7 games along total 825 experimental rounds (time, $t$). Time arrows indicate the duration of the games. (b) From left to right, matrix are the companion 2$\times$2 games to each of games~\cite{Binmore2001}. (c) Phase diagrams for the replicator dynamics of above companion 2$\times$2 payoff matrix, the theoretical expected direction of the social rotation.  Arrows in black  lines indicate the direction of trajectories of the evolution. Figure is made by the game dynamics program Dynamo \cite{Sandholm2011}. The positive (+) symbol indicate the expected evolutionary direction is counterclockwise, negative (-) is clockwise and zero (0) means no rotation. (d) The accumulate rotation $L^a$ of each of the 13 groups and initial by the 7 games respectively. The eigenvalue (calculation see text) for each game plotted in dash line in each game duration. Insert (right-up): Theoretical expected values $\lambda$ (includes the direction and the eigenvalue of the Jacobian) and the experimental mean rotation $\bar{L}$ (by rounds) values for the 5 real games in fulled dot, labeled with $\alpha$ (game ID).
}
\label{fig:B825matrix}
\end{figure}

This report starts from the Binmore, Swierzbinski and Proulx's experiment~\cite{Binmore2001}.
The experiment is repeated in 13 groups.  Each group contains 12 human subjects which are split into two populations to play random matching two-person companion 2$\times$2 games. Interesting is that, there are exists 7 different incentives (payoff matrix) along the 825 rounds game, or, the incentives parameter changes 6 times along the 825 periods (see Figure~\ref{fig:B825matrix}). This switching parameters experiment provides us a chance to insight (1) the changing direction and strength of social cyclic motions, and even more, (2) the inherence (relative dynamical response coefficient) of the social groups' dynamic property, quantitatively.

After introducing the measurement method and the experiment in section 2, in section 3 we show that (1) not only the direction but also the strength of the social rotation can be quantitatively distinguished; and (2) the diversity of the response coefficient cross the groups can be distinguished. In section 4, a potential theoretical explanation linking the strength of rotation to the payoff matrix is provided quantitatively, and then, discuss and summary last.

%

\section{2. Measurement and Data}

As a dynamics model,
EGT offers short run predictions where classic game theory has little to say.
Quantitative, in the 2$\times$2-like MSNE games~\cite{Binmore2001}, EGT predicts the evolution cyclic trajectory and rotation globally.
To capture the rotation, \emph{quantitatively}, is the main propose of this paper. The measurement for the rotation is introduced following.

\subsection{Strategy Space and Social Evolution}
For simple and clarity, we start from two-dimensional state space for
two-population random matching two-persons 2$\times$2 games. In EGT analogy~\cite{Sandholm2011}, social output can be described by the state space, the states and time. In the game, there exists two roles and each role includes some agents. For the first population $X$, the strategy set is $\{X_{1}$,$X_{2}\}$ for each agent; similarly, for the second population $Y$, $\{Y_{1}$,$Y_{2}\}$. Numerically, as ref.~\cite{selten2008}, we denote the strategy of $X_1$ (up strategy) and $Y_1$ (left strategy) as 1 following. The payoff matrix, as an example, is shown in Fig.~\ref{fig:B825matrix}(b).  For a given the payoff matrix, the solution of an evolutionary dynamics model can be seen in Fig.~\ref{fig:B825matrix}(c)~\cite{Sigmund1981} and two experimental trajectories can be seen in  Fig.~\ref{fig:SpaceVec}(b)~\cite{Binmore2001}. If there are $N$=6 agents in each population, an observable instantaneous populations strategy state should be  $x$:=$(p,q)\in \mathbb{X}$, herein $\mathbb{X}$ is the populations strategy state space and  $\mathbb{X}$=$\{0,\frac{1}{6},\frac{2}{6},\ldots,1\}  \times\{0,\frac{1}{6},\frac{2}{6},\ldots,1\}$, and $p$~($q$) is the density of $X_1$~($Y_1$) in $X$~($Y$). The strategy space is always an unit square.

Figure~\ref{fig:SpaceVec}(c) is an illustration for the state space of a 6+6 two-population 2$\times$2 game. The unit square $7 \times 7$ lattice (gray dots) is the state space  $\mathbb{X}$; In Fig.~\ref{fig:SpaceVec}(a), $A$ is the $x_A$=($\frac{5}{6},\frac{1}{6}$) is a state which means that $\frac{5}{6}$ in population $X$ choice $X_{1}$ meanwhile $\frac{1}{6}$ in population $Y$ choice $Y_{1}$.

In an experimental economics session, at each round, $t$, there is one dot of observation of $x(t)$ in $\mathbb{X}$. The vector, $x(t)$, is an indicator of collective social behavior at $t$. The \emph{smallest} tick $ \triangle t$ is $1$ and is the interval within the successive two rounds.
 Round by round, the dot ($x(t)$) should jump in the lattice and should form a trajectory.  Figure~\ref{fig:SpaceVec}(b) is an empirical example of a trajectory. The experimental system can be seen as a discrete-space and discrete-time population dynamics  system.

We investigate social dynamics in existed experimental data. Qualitatively, the authors~\cite{Binmore2001} confirm that, in the trajectories in some sessions,  as expected by evolutionary game theory or as Shapley (1964) cycle prediction, motion spiral pattern \emph{exists}. We go one step ahead. Quantitatively, we find the strength of the social  motion spiral can be distinguished and fit evolutionary game theory better than ever known.



 \subsection{Measurement for Rotation}

We are interested in the rotation observable -- a vector ($L$) called as rotation -- an indicator for the direction and the strength of the social evolutionary motion (or say as, the collective social dynamic behavior). In a two-population random matching two-persons 2$\times$2 game, $L$ can be seen as a one-dimensional vector perpendicular to the two-dimensional state space.

\begin{figure}
---------------------------------------------------------------------------------------------------------------------------------------------------------\\
\textbf{(a)}~~~~~~~~~~~~~~~~~~~~~~~~~~~~~~~~~~~~~~~~~~~~~~~~~~~~~~~~~~~~~~\textbf{(b)}~~~~~~~~~~~~~~~~~~~\\
---------------------------------------------------------------------------------------------------------------------------------------------------------\\
\centerline{
\includegraphics[width=5cm]{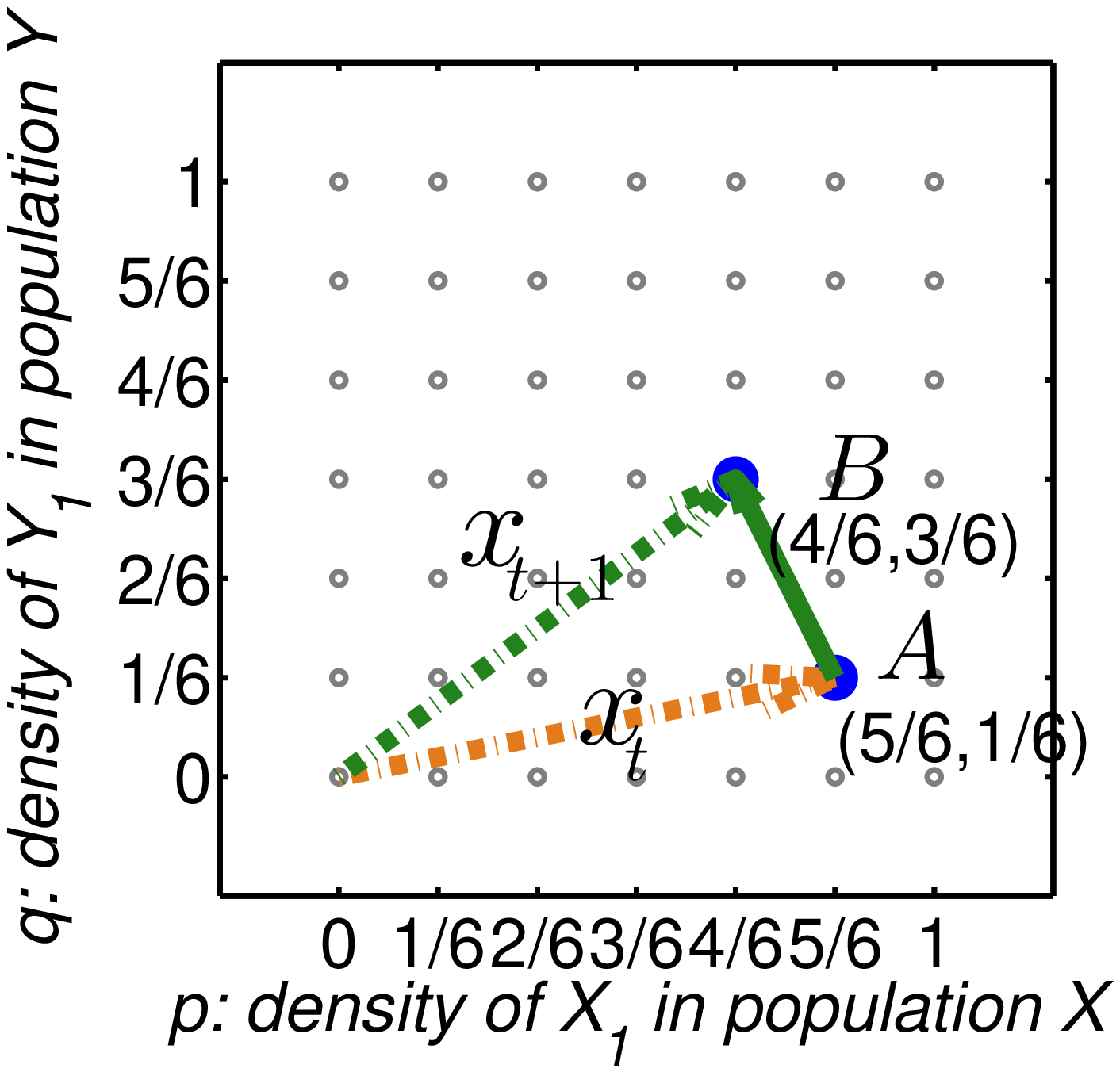}\includegraphics[width=5cm]{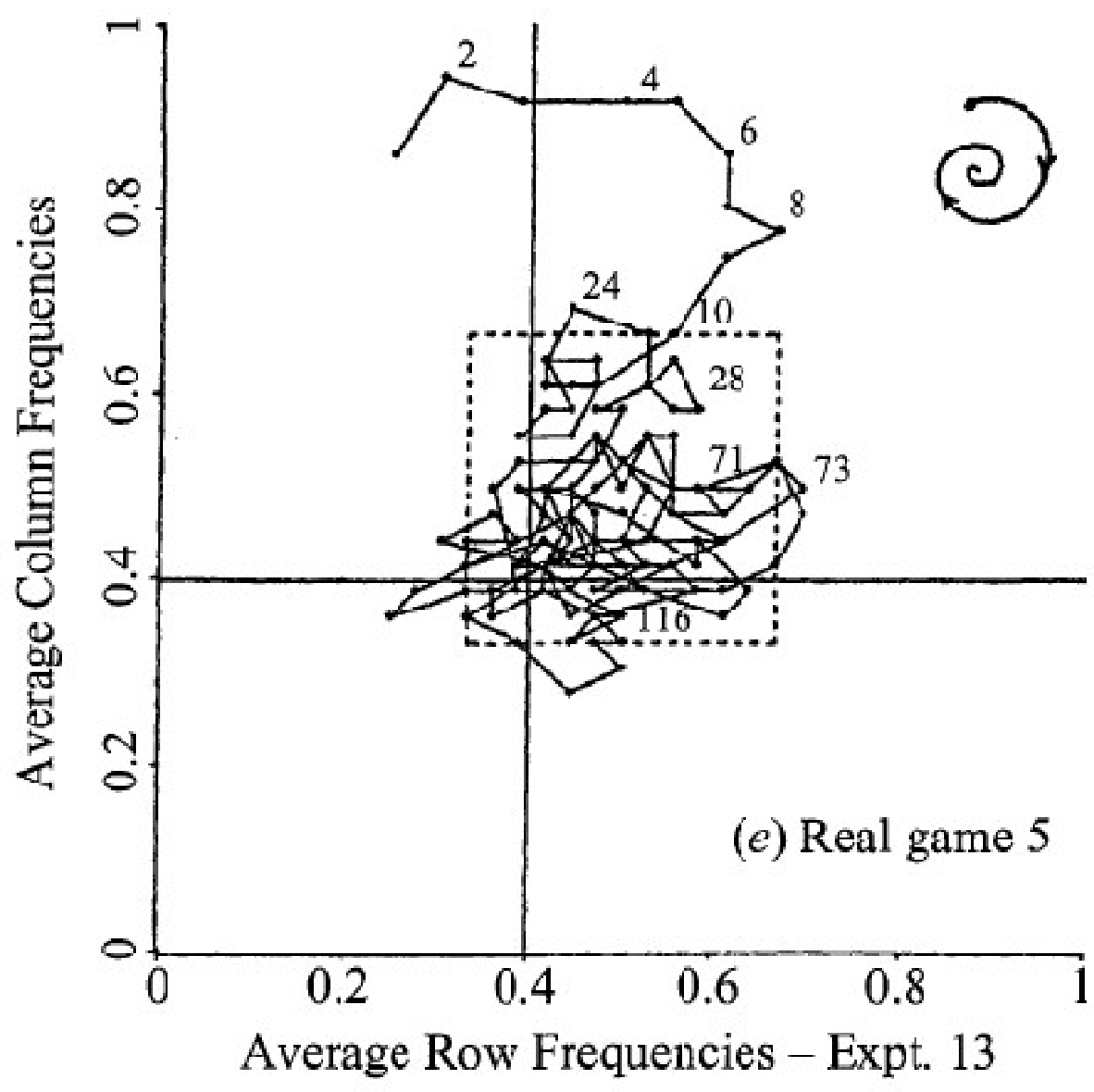} \includegraphics[width=5cm]{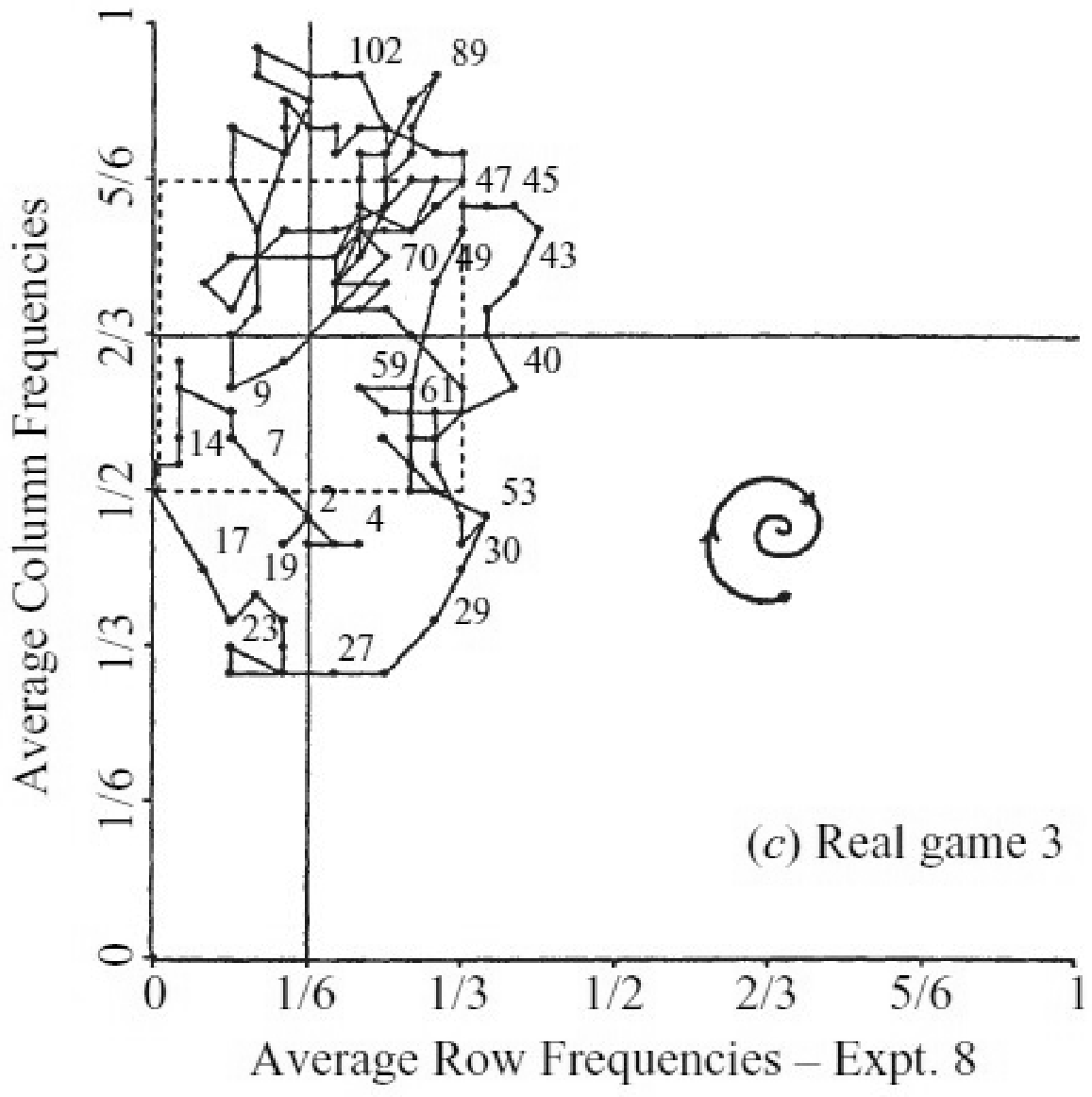} 
}
\caption{Measurement of the Social Rotation in State Space.
(a) Illustration of  an instantaneous social rotation $L$
 at time $t$. Supposing at time $t$, $x_t$ is at $A(5/6,1/6)$ and the next moment $t+1$, $x(t+1)$ is at $B(4/6,3/6)$;
  In this jumping from A to B, $L(t)$:=$x(t) \otimes x(t+1)$ is equal to 11/36. If a jumping was from B to A, its $L= -11/36$. Except the last round, every round has its observation of $L$.
(b) The trajectory of the social evolution in game 7 (or 5$th$ real game) and game 5 (or the 3$rd$ real game) in the experiment~\cite{Binmore2001}. The two figures are reproduced from ref.~\cite{Binmore2001}; Theoretically, spiral in left-hand-side should be clockwise and in right-hand-side should be counterclockwise.
}
\label{fig:SpaceVec}
\end{figure}

At experimental round $t$, the instantaneous rotation $L(t)$
can be expressed as
\begin{eqnarray}
    L(t) &=&  x(t) \otimes  x(t+1), 
\end{eqnarray}
where $x(t)$ is the strategy vector at time $t$, the $\otimes$ is cross product within the two strategy vectors.
 An example of an instantaneous rotation $L(t)$ is shown in Fig.~\ref{fig:SpaceVec}(a).
Denoting $T$ as the total number of  rounds  in game ($\alpha$) in group ($\beta$),
the accumulated $L$ (denoted as $L^a$) is
     $L_{\alpha,\beta}^a =  \sum_{t=0}^{T-1}\!L(t)$ in which $T$ is the total periods of the game $\alpha$ of the human subjects group $\beta$. The \emph{mean instantaneous rotation} can be denoted as $\bar{L}_{\alpha,\beta} = L_{\alpha,\beta}^a / T$.
A simplest picture is,  $\bar{L}_{\alpha,\beta}$ should depend on both $\alpha$ (an exogenous incentive of the experimental parameter setting) and $\beta$ (the inherence of a social group).  If $\bar{L}_{\alpha,\beta} > 0$, the rotation is counterclockwise; alternatively, clockwise.

We briefly explain the rotation observable $\bar{L}_{\alpha,\beta}$ here.
Given same $\alpha$, if $\bar{L}_{.,\beta} \neq \bar{L}_{.,\beta'}$, we can say, the "social material" $\beta$ is unequal to $\beta'$, or say, the response coefficients of the materials differ. Meanwhile, given same $\beta$,  changing $\alpha$ is changing incentive, and then, the strength of rotation should be changed.
In statistical physics language, $\bar{L}$ relates to probability current or flux in state space. This observable relates Shapley polygon~\cite{Shapley1964} and stability of social dynamics (e.g., \cite{Cason2010}).
This observable can also be imaged as angular momentum, one of the most fundamental variable in physics, but set mass term as 1.


%

\subsection{Data from Switching Parameter Experiments }

The controlled parameters in experimental games are of the payoff matrix. The original payoff matrix can be seen in Figure~\ref{fig:B825matrix}(a) in which the row players's payoff is given; the column player's payoff is its opposite value; all of the games are of zero-sum game. Figure~\ref{fig:B825matrix}(b) are the zero-sum payoff matrix of the companion 2$\times$2 games in Figure~\ref{fig:B825matrix}(a)~\cite{Binmore2001}.

The random matching two-persons pair comes from the two populations respectively.
The total length of a groups playing is of 825 rounds,
with 13 groups containing  12 human subjects.  The 12 human subjects are split into two-population equally in which one population play $X$-role and another population play $Y$-role.  Each group plays all 7 games in the 825 rounds. The Game 1 and Game 2 are practice zero-sum game and not real game are provided to the human subjects for training.
All of the 156 subjects comes from student subject pool in University of Michigan, U.S. The experiments are conducted in the end of 1990s. For more detail of the parameters of the 7 experimental games, see ref.~\cite{Binmore2001}

\section{results}

\subsection{Gross Figural and Numerical Results }

Figure~\ref{fig:B825matrix}(d), the 13 color curves are the empirical accumulate rotation $L^a$ of each of the 13 groups and initial by the 7 games respectively. The successive increasing or declining of the $L^a$ means the successively counterclockwise or clockwise rotation is observed. Numerical results see Table~\ref{tab:AccumulateR}.  In these data,   (1) in the game dimension, we explore the consequence of the switching the payoff matrix; (2) in the human subject groups dimension, we explore the inherence response coefficient cross the groups.

\begin{table}[htbp]
\caption{Empirical Accumulate Rotation ($L^a$) of the 13 Groups in the 7 Games}
\label{tab:AccumulateR}
\centering
\begin{tabular}{|	c|	r	r	r	r	r	r	r	r	r	rr	r r|rr|}
  \hline 				
  Group$(\beta)$	&1	&2	&3	&4	&5	&6	&7	&8	&9	&10	&11	&12	&13 &&	 	\\
Game($\alpha$)	&T61	&T62	&T63	&T64	&T65	&T66	&T67	&T68	&T70	&T71	&T72	&T73	&T74	 &Avg.	 &S.E.	\\
  \hline 				
1	&~~-0.11	&~~0.11	&~~0.03	&~~0.06	&~~-0.14	&~~0.14	&~~-0.44	&~~0.11	&~~0.00	&~~0.03	&~~-0.19	&~~0.03	 &~~-0.17	&~~-0.04	&~~0.16	\\
2	&~~-1.69	&~~-0.81	&~~-0.36	&~~-0.56	&~~-0.33	&~~-0.5	&~~-0.33	&~~-0.56	&~~-0.92	&~~-0.14	 &~~-0.03	&~~0.11	&~~0.17	&~~-0.46	&~~0.49	\\
3	&~~0.89	&~~0.67	&~~1.47	&~~1.89	&~~0.78	&~~0.81	&~~1.17	&~~0.39	&~~2.53	&~~0.72	&~~0.97	&~~0.72	&~~0.89	&~~1.07	 &~~0.58	 \\
4	&~~0.39	&~~-0.42	&~~0.14	&~~-0.17	&~~0.56	&~~0.14	&~~0.36	&~~-0.39	&~~0.19	&~~-0.22	&~~0.17	&~~0.17	 &~~0.03	 &~~0.07	&~~0.3	\\
5	&~~1.25	&~~0.86	&~~1.03	&~~0.94	&~~1.11	&~~1.14	&~~1.58	&~~0.53	&~~1.39	&~~1.72	&~~0.44	&~~1.31	&~~0.61	&~~1.07	 &~~0.39	 \\
6	&~~-1.47	&~~-0.92	&~~-2.33	&~~-1.5	&~~-1.06	&~~-1.25	&~~-1.53	&~~-2.5	&~~-2.75	&~~-3.06	 &~~-2.03	&~~-2.42	&~~-2.69	&~~-1.96	&~~0.71	\\
7	&~~-3.00	&~~-1.31	&~~-1.92	&~~-2.97	&~~-0.94	&~~-2.61	&~~-1.75	&~~-1.36	&~~-2.67	&~~-4.33	 &~~-2.47	&~~-1.78	&~~-2.47	&~~-2.28	&~~0.9	\\
\hline
R.R.C.	&~~1.02	&~~0.62	&~~1.09	&~~1.18	&~~0.68	&~~0.9	&~~1.03	&~~0.68	&~~1.56	&~~1.44	&~~0.86	&~~0.98	&~~0.97	&~~1	 &~~0.28	
\\
\hline
\end{tabular}
\begin{flushleft}
R.R.C. is the abbreviation of relative response coefficient ($\Phi_{.,\beta}$) of each of the 13 human subject groups. The last row is the mean $\Phi_{.,\beta}$ cross the game $\alpha=3,5,6,7$. The second row (e.g., T61 to T74) is the labeled according to the session order of  original experimental record in which there is not session labeled as 69.
The mean instantaneous rotation $\bar{L}=L^a/T$ in which $T$ is the total round of an experimental session of a given payoff matrix.
\end{flushleft}
\end{table}

\begin{figure}[!t]
---------------------------------------------------------------------------------------------------------------------------------------------------------\\
~~~~~~~~~~~~~~~~~~~\textbf{(a)}~~~~~~~~~~~~~~~~~~~~~~~~~~~~~~~~~~~~~~~~~~~~~~~~~~~~~~~~~~~~~~~~~~~~\textbf{(b)}~~~~~~~~~~~~~~~~~~~\\
---------------------------------------------------------------------------------------------------------------------------------------------------------\\

\centerline{\includegraphics[width=0.40\textwidth]{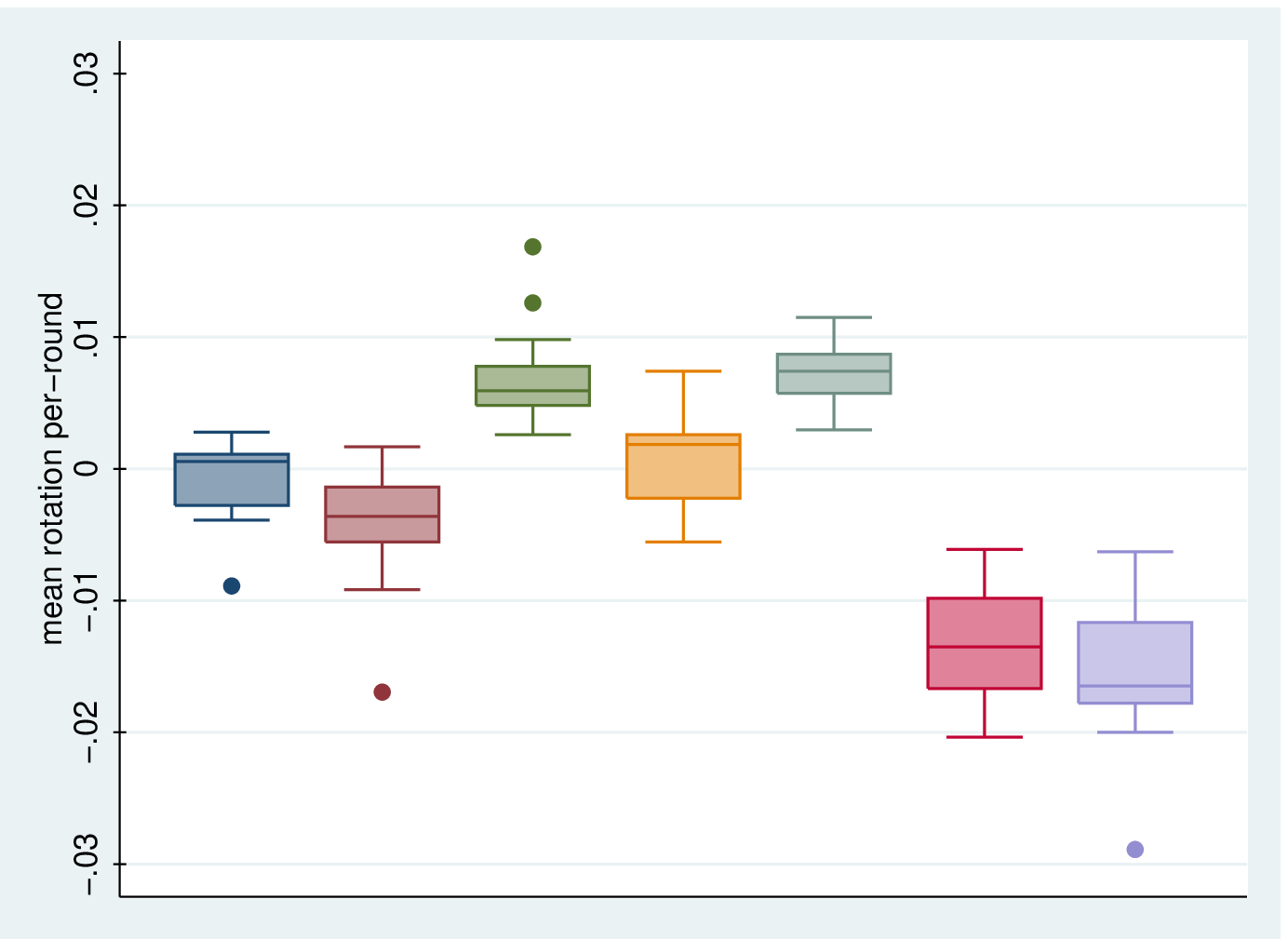}~~~~{\includegraphics[width=0.40\textwidth]{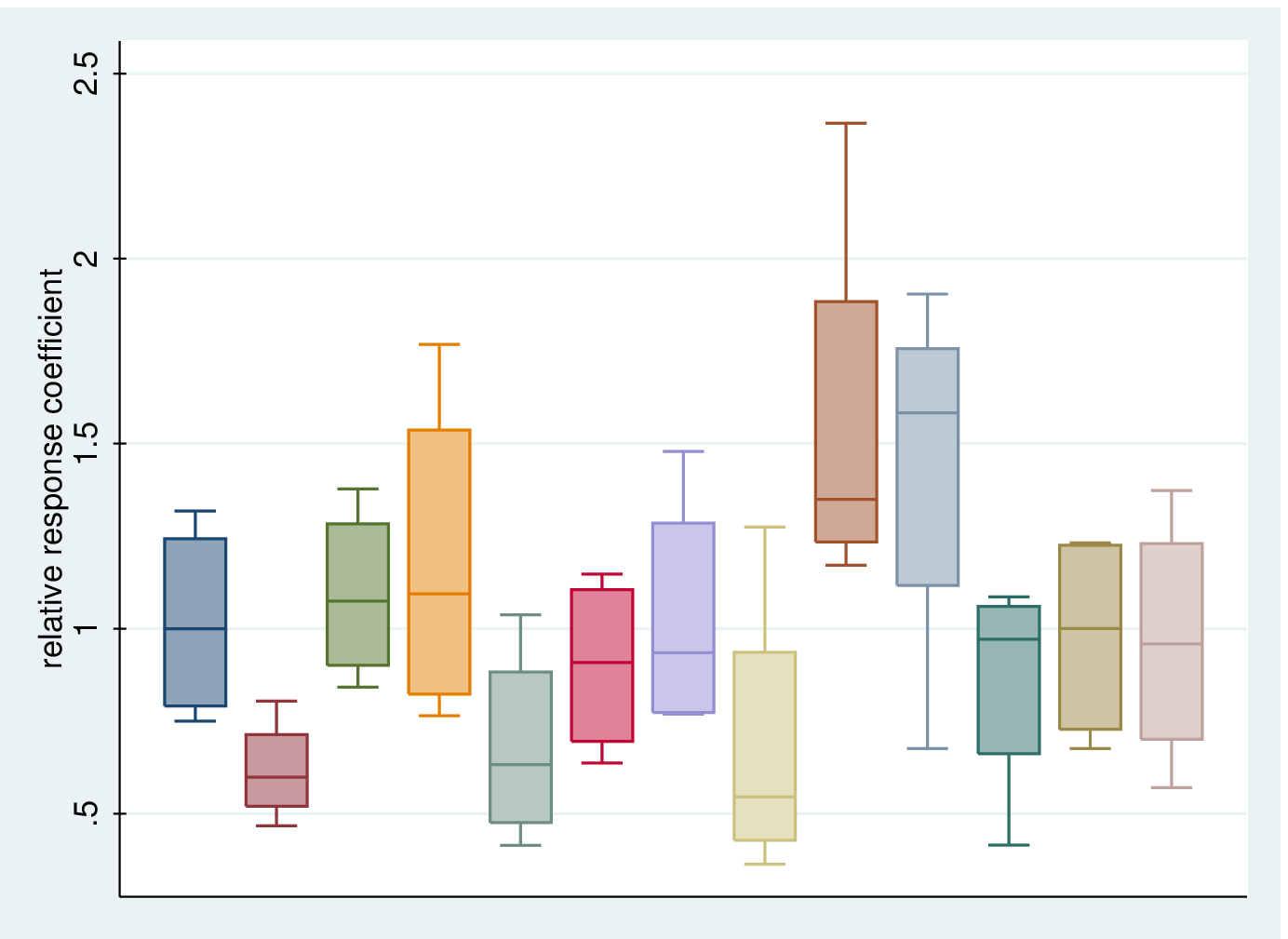}}}
\caption{Empirical results. (a) Figural presentation for Result-1.1 and Result-1.2. The  box-plot of the mean rotation $\bar{L}_{\alpha,.}$ by Games ($\alpha=1,2,...,7$), in which Groups ($\beta=1,2,...,13$) results are pooled by the 7 games (Result 1.1 and 1.2 in text). (b) Figural presentation for Result-2.1 and Result-2.2. Empirical Result by Groups. The box-plot of the relatively strength $\Phi_{.,\beta}$. From left to right are Group 61-68 and Group 70-74. Each group $\beta$ has 4 samples from Game 3, 5, 6 and 7. A box is the 4 values' range of $\Phi_{\alpha,.}$ for a given $\beta$ (From left to right, $\beta=1,2,...,13$).}
\label{fig:boxplot7game}
\end{figure}

%

%

\subsection{Cross the 4 MSNE Games: Rotation Strength and Incentive}

Theoretically, payoff matrix relates to the incentive of the individual in an experimental social. It is straight to image that,  payoff matrix could be links to the strength of social rotation. Different game with different payoff matrix could have different strength of social rotation.

Numerical results of the accumulated rotation of the 13 Groups in the 7 Games are shown in Table~\ref{tab:AccumulateR}.  Figure~\ref{fig:boxplot7game}(a) are the figural results. Both show the direction (positive is along counter-clock) and the strength  of the $L^a_{\alpha,\beta}$, respectively. From Table~\ref{tab:AccumulateR}, row by row's or game by game's statistical analysis comes to following two results.

\textbf{Result-1.1: }Numerical analysis show that, the direction in the four games (each with 13 samples) are consistence with evolutionary game theory expectations, respectively ($p < 0.00$). None of the 52 samples deviate the theoretical prediction.

\textbf{Result-1.2:} Numerical analysis also show that, the absolutely value of the strength cross the 4 MSNE games can be distinguished. Disregards the direction,  numerical results shows that, rotation strength values of both of the Game 6 and Game 7 significant larger than those of the Game 3 and Game 5 ($p < 0.01, t$-$test$).

A potential interpretation for the strengths cross the games is given in next section.


\subsection{Cross the 13 Human Subjects Groups: Diversity of Response Coefficient }

For individual, the sensitive on the incentive could be different and can be detected in game~\cite{Hsu2007,XuWang2011EPR}; For group, the sensitives are different and detectable~\cite{Camerer2003}. The diversity of response character be obtained from the observable -- the strength of social rotation.

Within Groups, there is 4 games taken into account to test the diversity on response. We do not include the Game 4, for it is not the MSNE, the observable of rotation is trivial. From Table~\ref{tab:AccumulateR}, column by column's (group by group's) statistical analysis comes to the results: Result-2.1 and Result-2.2.

\textbf{Result-2.1:}
Result-2.1: In the direction of rotation view, none of the 52 (13 group and the 4 games) observed $L^a$ deviates the
prediction from the evolutionary game theory. The prediction of the replicator dynamics (the most simple EGT model) for the 4 games are shown in Figure~\ref{fig:B825matrix}(c).

None of the groups deviating from EGT's prediction. This means, in the 12 subjects systems, the social rotation (and also Shapley-cyclic motion) directions obey the EGT. In this view, the similarity between the groups is significant.

\textbf{Result-2.2:} In the sensitive of the groups response in rotation view, from  Table~\ref{tab:AccumulateR},   Figure~\ref{fig:boxplot7game}(b) are the figural results for the relative accumulated rotation $\Phi_{\alpha,\beta}$ ($\Phi_{\alpha,\beta} = L_{\alpha,\beta}^a / L'_{\alpha}$, here, $L'_{\alpha}=\sum_{\beta}\frac{L_{\alpha,\beta}^a}{S_\beta}$ and sample size $S_\beta=13$ in our case) cross the 13 groups and each group has 4 ($\alpha=3,5,6,7$) samples, respectively. That is to say, for given $\beta$, if $\Phi_{\alpha,\beta}$ is strictly larger than 1, this suggests that, the sample $\beta$ is more sensitive. $\Phi_{\alpha,\beta}$ can be a quantity relating to response coefficient of group $\beta$. Cross the games ($\alpha=3,5,6,7$), the relative accumulated rotation $\Phi_{.,\beta}$ of $\beta$-group is shown in last row Table~\ref{tab:AccumulateR}.

We use Kruskal-Wallis equality-of-populations rank test method to evaluate the groups' response coefficient. Each of the 13 groups is used as a group sample; in each group sample, there is 4 MSNE games's relative rotation strength observation.

Result shows that, the equality hypothesis can be rejected ( $p<0.05$,  $\chi^2$=21.330 and 12 degrees of freedom).
This result means the different is not random result of the 13 samples. The groups' \emph{inherence} (response coefficient) is different and could be persisted cross the games.
%
%

%
%
%
%
%
%

Some samples results are: the group 62, 65 and 68 is less ($\Phi_{\alpha,\beta} <$ 1 at $p<0.1$ significant) sensitive to the incentives; Alternatively, the group 70 is the most sensitive ($\Phi_{\alpha,\beta} >$ 1 at $p<0.01$ significant) to the payoff matrix switching.
The numeric results of the relative response coefficients (R.R.C.) can be read from the last row in Table~\ref{tab:AccumulateR}.

\section{A Theoretical Explanation for the Rotation}
Trying to find out the theoretical explantation of the strength, we start from the simplest evolutionary dynamical equation -- replicate dynamics equation specified by the 2$\times$2 payoff matrix for the two population games. Symbols in this section is following reference~\cite{Sigmund1981}.

\textbf{On the direction of rotation}. Figure~\ref{fig:B825matrix}(c) provides all 7 games pattern with the Dynamo~\cite{Sandholm2011}. In direction view, all of the results from real experiment (in Game 3, 5, 6 and 7) total 52 sessions meet the theoretical model exactly ($p<0.00$). In Game 4, cross the 13 sample, 13 samples statistically not deviate from zero hypothesis ($t$-$test>$0.1) which is expected by EGT, because, in these systems, no rotation exists.

\textbf{On the strength of rotation}. Theoretically, eigenvalues of the payoff matrix determine the dynamics property like convergence and stability (e.g.,~\cite{Selten1991Tangused}).
This point is supported in experiments (e.g.,~\cite{Tang2001}).
Each 2$\times$2 payoff bi-matrix can be simplified to the two matrix
\begin{center}
$\left(
  \begin{array}{cc}
    0 & a \\
    b & 0 \\
  \end{array}
\right)$ and $\left(
  \begin{array}{cc}
    0 & c \\
    d & 0 \\
  \end{array}
\right)$
\end{center}
as reference~\cite{Sigmund1981}. Then, the two matrix are \emph{multiply} to satisfied the concentrations:
 \begin{equation}\label{eq:abcd1}
     |a+b|=1~~\cap~~|c+d|=1,
 \end{equation}
And the efficient payoff matrix can be obtained. As mentioned in~\cite{Sigmund1981}, there exists  an unique equilibrium ($|c|, |a|$) in the interior space  if   and   only   if  $ab >  0 $ and $cd> 0$  (Game 2, 3, 5, 6 and 7 are in this condition). If not this condition, the dynamic equation is trivial (Game 1 and 4, the systems are not  oscillating but converging to pure Nash states). In $ac < 0$ case, what is interested in this report, the eigenvalues $v$ of the Jacobian of the dynamic equation evaluated near the unique equilibrium point can be expressed as~\cite{Sigmund1981},
\begin{equation}\label{vv}
     v = \pm i  \lambda = \pm i \sqrt {abcd}.
\end{equation}

%

 Table~\ref{tab:abcdSigmundList} is the results along above steps and shows the efficient payoff matrix elements ($a, b, c, d$). Table~\ref{tab:abcdSigmundList} also shows the eigenvalues ($\lambda$) from Jacobian of replicator dynamics model of the given games.
 It is well known that the eigenvalue $\lambda$ direct proportion to the frequency (or angular velocity) of vibration; meanwhile, empirically~\cite{WangInertia2012}, mean $L$  proportion to mean angular velocity in MSNE systems. So our theoretical expectation is:
\begin{center}
 $\bar{L}_{\alpha,.} \propto \lambda_\alpha$.
\end{center}
 Significantly, the rank of the theoretical predicted strength $\bar{L}$ cross the games fit the games \emph{exactly}. For visible, the theoretical values and the experimental mean rotation values for the games are shown in Figure~\ref{fig:B825matrix}(d)-Insert subplot. The linear relationship --- $\bar{L}_\alpha$ vs $\lambda_\alpha$  --- looks significant.

%
 Before the end of this section, we have to emphasis that, the constrained payoff method suggested in Eq.~\ref{eq:abcd1} is only a suggestion which needs more empirical testing, even though  the results from the theoreital suggestion and results from experimental data meet well.


\begin{table}[htbp]
\caption{Efficient Payoff Matrix Elements and the Eigenvalues}
\label{tab:abcdSigmundList}
\centering
\begin{tabular}{r c	r	r	r	r	r 	c}
  \hline 			%
& Game($\alpha$)	&~~~~~$a$~~~	&~~~~~$b$~~~	&~~~~~$c$~~~	&~~~~~$d$~~~	& ~~~~~$|\lambda|$~~~	& ~~~~~~Nash~~~~~~ \\
  \hline 			%
& 1	&-1	&2	&-2	&1	& --	& (0, 0)	\\
& 2	&-1/3	&-2/3	&1/3	&2/3	&0.222	& (1/3, 1/3)	\\
& 3	&5/6	&1/6	&-1/6	&-5/6	&0.139		& (1/6, 5/6) \\ 
& 4	&4	&-3	&2	&-3	& --		&  (1, 1) \\
& 5	&2/3	&1/3	&-1/6	&-5/6	&0.176 	& (1/6, 2/3)	\\
& 6	&-1/3	&-2/3	&1/3	&2/3	&0.222		& (1/3, 1/3)\\
& 7	&-2/5	&-3/5	&2/5	&3/5	&0.240		& (2/5, 2/5)\\
\hline
\end{tabular}
\begin{flushleft}
$|\lambda|$ is the solution of the Jacobian of the constrainted replicator dynamics differential equation, e.g., ref~\cite{Sigmund1981}. Both of the 2-nd strategies for the Game ($\alpha=1$) are the 3-rd strtegies; The two games  ($\alpha=1,4$)  are not MSNE, the Nash are evaluated from the companion 2$\times$2 payoff-matrix directly.
\end{flushleft}
\end{table}


%
%


\section{Discussion}

%

In this section, we explain the potential application of our finding for future experimental 
and theoretical investigations. Then we compare our finding with two recently experimental reports.




(1)\textbf{ Application to  experimental investigation} --- Evolutionary game theory usually analysis mathematically a continuous time hypothesis, however, experimental system usually not. Evaluating the dynamics in a discrete-space and discrete-time experimental data is an approximation. Continuous strategy in population can provide more information. Meanwhile, the scale of time is also a dimension needed to be considered. The continue time game system has firstly  been developed by Daniel Friedman group and is providing deeper understanding in game behavior~\cite{Friedman2011}, and the technical development should provide a deeper experimental economics insight for EGT.

Analysis social dynamical performance on the full state space lattice emerges recently; The evolution velocity pattern~\cite{XuWang2011ICCS}, the evolution stability~\cite{nowak2012}
 have been qualitatively distinguished in experiments.
 Meanwhile, the performance of social in game has been linked to fluctuation theorem in statistical physics~\cite{XuWang2012LDF,XWMm2011} and fundamental principle (e.g., maximum entropy principle~\cite{xuetal2012Maxent}) in nature science. 
 All these regularities are found in full social strategy space. We suggest that
testing observable in full space could provide much more dynamics information form experiment data.

%



(2) \textbf{Application to theoretical investigation} ---
There are variety of
plausible dynamics which describe adaptive mechanisms underlying game theory. The result from the metrics of rotation can provides a novel way to evaluate theory.
%
%
Applying to test existed theorem, Minimax randomization hypothesis can be strictly rejected by the observations of rotation. Randomization hypothesis state that the player in MSNE (Game 3, 5, 6 and 7) should
play full randomly, so its prediction on rotation is strictly zero (zero-Hypothesis). The observations of rotation in all of the 52 experimental sessions strictly deviate from the zero-Hypothesis (at $p<0.000$).

 It is no surprise that, if quantitative observable can be indexed more efficiently, the co-play of evolutionary game theory and experimental economics can continue. For example, the mean rotation $\bar{L}$ as an empirical observable and the velocity field (method see reference~\cite{XuWang2011ICCS}) as another empirical observable, both of which can be used to evaluating variety evolution dynamics.

(3) \textbf{Comparison with the two experimental reports~\footnote{This paragraph is added to the version arxiv:1203.2591 after we receive the reference~\cite{Friedman2012}. We thank the authors to send us this critical reference timely. }} --- Recently, two experimental reports provides empirical supporting on evolution dynamics positively which are the most relative to our report.

The first paper is reported by Hoffman, Sieget, Nowak and Greenz (HSNG)~\cite{nowak2012}; They test the distribution differences in three variety parameters games; In their experimental setting, one of the trajectory should be spiral in (Good-rps), one should be in middle (Neutral-rps) and one should be spiral out (Bad-rps).  They report the Bad-rps differ from the others two and  meet EGT prediction well. The difference within Good-rps and Neutral-rps can not be distinguish in their method. The main index method is called as \emph{average distance} by measure the average Euclidean distance of social state from their theoretical equilibrium ($\frac{1}{3}, \frac{1}{3}, \frac{1}{3}$) by periods.

The second paper is the report from Cason, Friedman and Hopkins (CFH)~\cite{Friedman2012}. The authors use two method to distinguish the
evolution pattern. The first method is the same as the \emph{average distance} as HSNG (called as \emph{cycle amplitude} in CFH paper); Using this observable from experiments, the authors excludes the TASP model and illustrate also that the parameters of the perturbed best response dynamics can be estimated.

Interesting is the second method --- \emph{cycle rotation index} ($CRI$) in CFH method. To calculate this index, a
line segment is constructed between the Nash equilibrium and the simplex edge of the strategy space of the game; The segment serves as a "tripwire" for counting cycles. In the stochastic processes, the social particle should passes the segment from left or right, called as counter-clockwise transits ($CCT$) or clockwise ($CT$). The cycle rotation index
then is defined for each period as $CRI = (CCT-CT)/(CCT+CT) \in [-1, 1]$. In this way, in quite wide parameters, the existence of the cycles can be distinguished straightly.

Now we analysis the similarity and difference of the two papers and ours (WX) reported here. All these three reports are for testing EGT. None of them rejects the predictions EGT.  As point out by CFH, no rotation reported in HSNG and so the directions of the rotation can not found in HSNG, meanwhile, both of CFH and WX report the directions of social rotations meet EGT. Furthermore, both of CFH and WX reports relates to the rotation quantitatively. CFH's $CRI$ method is more visible and less abstract, meanwhile, WX's $\bar{L}$ method has its fundament in  classical physics and has its mathematical simplicity.

in summary, better measurements of the dynamical observable should improve our understanding and developing of EGT; Meanwhile, theoretical investigations are also more expected while the dynamical pattern are being seen in CFH, HSNG and ours.

\section{Summary}


Main findings from the zero-sum experimental games are (1) Shapley-cycles existence indeed in all of the MSNE systems. The rotation directions are governed by evolutionary game theory. None of real game sample violates.
 (2) The strength of rotation could be captured by evolutionary game theory quantitatively;
  (3) The dynamics inherence of a group can persist cross the switching incentive parameter games.
All above results come from the observation --- evolutionary rotation vector $L$ --- in the switching incentive zero-sum games~\cite{Binmore2001}.

These results can be understood as:
The rotations of a motor are governed by the exogenous electric power provided; At the same time, different motor has its own inherence dynamic property respectively. Similarly, our finding state that, the laboratory social rotation can be controlled by the payoff matrix; meanwhile, each \emph{human subject social motor} could has its own inherence response coefficients when inner group individual interactions exist.

In the book, von Neumann and Morgenstern (1944) state that \emph{We repeat most emphatically that our theory is thoroughly static. A dynamic theory would unquestionably be more complete and preferable.} As emphasis by Tang,\emph{ theoretical constructs must be built based on robust empirical findings}~\cite{Tang2010}, we wish the robust empirical dynamical quantitative patterns reported in this letter, companion with the dynamical pattern found in variety games~\cite{XuWang2011,XuWang2011EPR}, could provide useful information and constraint for future dynamic theory of game.

We suggest that, Shapley-cycles can be more quantitatively understood, including the shapes, the directions and the strengths, in more general game environments. \\

\newpage
\textbf{Acknowledgement}: We are grateful to Ken Binmore's patient discussion, critical command and encourage on our work in this direction. We thank Joe Swierzbinski provides our data and answer our questions in huge patient. Technology support provided by Zunfeng Wang is thanks.

%


%
%


\begin{thebibliography}{25}
\expandafter\ifx\csname natexlab\endcsname\relax\def\natexlab#1{#1}\fi
\expandafter\ifx\csname bibnamefont\endcsname\relax
  \def\bibnamefont#1{#1}\fi
\expandafter\ifx\csname bibfnamefont\endcsname\relax
  \def\bibfnamefont#1{#1}\fi
\expandafter\ifx\csname citenamefont\endcsname\relax
  \def\citenamefont#1{#1}\fi
\expandafter\ifx\csname url\endcsname\relax
  \def\url#1{\texttt{#1}}\fi
\expandafter\ifx\csname urlprefix\endcsname\relax\def\urlprefix{URL }\fi
\providecommand{\bibinfo}[2]{#2}
\providecommand{\eprint}[2][]{\url{#2}}

\bibitem[{\citenamefont{Binmore et~al.}(2001)\citenamefont{Binmore,
  Swierzbinski, and Proulx}}]{Binmore2001}
\bibinfo{author}{\bibfnamefont{K.}~\bibnamefont{Binmore}},
  \bibinfo{author}{\bibfnamefont{J.}~\bibnamefont{Swierzbinski}},
  \bibnamefont{and} \bibinfo{author}{\bibfnamefont{C.}~\bibnamefont{Proulx}},
  \bibinfo{journal}{The Economic Journal} \textbf{\bibinfo{volume}{111}},
  \bibinfo{pages}{445} (\bibinfo{year}{2001}).

\bibitem[{\citenamefont{Bena{\=\i}m et~al.}(2009)\citenamefont{Bena{\=\i}m,
  Hofbauer, and Hopkins}}]{benaim2009learning}
\bibinfo{author}{\bibfnamefont{M.}~\bibnamefont{Bena{\=\i}m}},
  \bibinfo{author}{\bibfnamefont{J.}~\bibnamefont{Hofbauer}}, \bibnamefont{and}
  \bibinfo{author}{\bibfnamefont{E.}~\bibnamefont{Hopkins}},
  \bibinfo{journal}{Journal of Economic Theory} \textbf{\bibinfo{volume}{144}},
  \bibinfo{pages}{1694} (\bibinfo{year}{2009}), ISSN \bibinfo{issn}{0022-0531}.

\bibitem[{\citenamefont{Gaunersdorfer and Hofbauer}(1995)}]{Hofbauer1995}
\bibinfo{author}{\bibfnamefont{A.}~\bibnamefont{Gaunersdorfer}}
  \bibnamefont{and} \bibinfo{author}{\bibfnamefont{J.}~\bibnamefont{Hofbauer}},
  \bibinfo{journal}{Games and Economic Behavior} \textbf{\bibinfo{volume}{11}},
  \bibinfo{pages}{279} (\bibinfo{year}{1995}).

\bibitem[{\citenamefont{Cason et~al.}(2010)\citenamefont{Cason, Friedman, and
  Hopkins}}]{Cason2010}
\bibinfo{author}{\bibfnamefont{T.}~\bibnamefont{Cason}},
  \bibinfo{author}{\bibfnamefont{D.}~\bibnamefont{Friedman}}, \bibnamefont{and}
  \bibinfo{author}{\bibfnamefont{E.}~\bibnamefont{Hopkins}},
  \bibinfo{journal}{Journal of Economic Theory} \textbf{\bibinfo{volume}{145}},
  \bibinfo{pages}{2309} (\bibinfo{year}{2010}).

\bibitem[{\citenamefont{Falk and Heckman}(2009)}]{Falk2009}
\bibinfo{author}{\bibfnamefont{A.}~\bibnamefont{Falk}} \bibnamefont{and}
  \bibinfo{author}{\bibfnamefont{J.}~\bibnamefont{Heckman}},
  \bibinfo{journal}{Science} \textbf{\bibinfo{volume}{326}},
  \bibinfo{pages}{535} (\bibinfo{year}{2009}).

\bibitem[{\citenamefont{Samuelson}(2002)}]{Samuelson2002}
\bibinfo{author}{\bibfnamefont{L.}~\bibnamefont{Samuelson}},
  \bibinfo{journal}{The Journal of Economic Perspectives}
  \textbf{\bibinfo{volume}{16}}, \bibinfo{pages}{47} (\bibinfo{year}{2002}).

\bibitem[{\citenamefont{Sandholm}(2011)}]{Sandholm2011}
\bibinfo{author}{\bibfnamefont{W.}~\bibnamefont{Sandholm}},
  \emph{\bibinfo{title}{Population games and evolutionary dynamics}}
  (\bibinfo{publisher}{MIT press Cambridge, MA:}, \bibinfo{year}{2011}).

\bibitem[{\citenamefont{Selten and Chmura}(2008)}]{selten2008}
\bibinfo{author}{\bibfnamefont{R.}~\bibnamefont{Selten}} \bibnamefont{and}
  \bibinfo{author}{\bibfnamefont{T.}~\bibnamefont{Chmura}},
  \bibinfo{journal}{The American Economic Review}
  \textbf{\bibinfo{volume}{98}}, \bibinfo{pages}{938} (\bibinfo{year}{2008}).

\bibitem[{\citenamefont{Schuster and Sigmund}(1981)}]{Sigmund1981}
\bibinfo{author}{\bibfnamefont{P.}~\bibnamefont{Schuster}} \bibnamefont{and}
  \bibinfo{author}{\bibfnamefont{K.}~\bibnamefont{Sigmund}},
  \bibinfo{journal}{Animal Behaviour} \textbf{\bibinfo{volume}{29}},
  \bibinfo{pages}{186} (\bibinfo{year}{1981}).

\bibitem[{\citenamefont{Shapley}(1964)}]{Shapley1964}
\bibinfo{author}{\bibfnamefont{L.}~\bibnamefont{Shapley}},
  \bibinfo{journal}{Advances in game theory} \textbf{\bibinfo{volume}{52}},
  \bibinfo{pages}{1¨C29} (\bibinfo{year}{1964}).

\bibitem[{\citenamefont{Hsu et~al.}(2007)\citenamefont{Hsu, Huang, and
  Tang}}]{Hsu2007}
\bibinfo{author}{\bibfnamefont{S.}~\bibnamefont{Hsu}},
  \bibinfo{author}{\bibfnamefont{C.}~\bibnamefont{Huang}}, \bibnamefont{and}
  \bibinfo{author}{\bibfnamefont{C.}~\bibnamefont{Tang}}, \bibinfo{journal}{The
  American Economic Review} \textbf{\bibinfo{volume}{97}}, \bibinfo{pages}{517}
  (\bibinfo{year}{2007}).

\bibitem[{\citenamefont{Xu and Wang}(2011{\natexlab{a}})}]{XuWang2011EPR}
\bibinfo{author}{\bibfnamefont{B.}~\bibnamefont{Xu}} \bibnamefont{and}
  \bibinfo{author}{\bibfnamefont{Z.}~\bibnamefont{Wang}},
  \bibinfo{journal}{Arxiv preprint arXiv:1107.6043}
  (\bibinfo{year}{2011}{\natexlab{a}}).

\bibitem[{\citenamefont{Camerer and Foundation}(2003)}]{Camerer2003}
\bibinfo{author}{\bibfnamefont{C.}~\bibnamefont{Camerer}} \bibnamefont{and}
  \bibinfo{author}{\bibfnamefont{R.~S.} \bibnamefont{Foundation}},
  \emph{\bibinfo{title}{Behavioral game theory: Experiments in strategic
  interaction}}, vol.~\bibinfo{volume}{9} (\bibinfo{publisher}{Princeton
  University Press Princeton, NJ}, \bibinfo{year}{2003}).

\bibitem[{\citenamefont{Selten}(1991)}]{Selten1991Tangused}
\bibinfo{author}{\bibfnamefont{R.}~\bibnamefont{Selten}},
  \textbf{\bibinfo{volume}{1}}, \bibinfo{pages}{98?54} (\bibinfo{year}{1991}).

\bibitem[{\citenamefont{Tang}(2001)}]{Tang2001}
\bibinfo{author}{\bibfnamefont{F.}~\bibnamefont{Tang}},
  \bibinfo{journal}{Journal of economic behavior $\&$ organization}
  \textbf{\bibinfo{volume}{44}}, \bibinfo{pages}{221} (\bibinfo{year}{2001}).

\bibitem[{\citenamefont{Xu and Wang}(2012{\natexlab{a}})}]{WangInertia2012}
\bibinfo{author}{\bibfnamefont{B.}~\bibnamefont{Xu}} \bibnamefont{and}
  \bibinfo{author}{\bibfnamefont{Z.}~\bibnamefont{Wang}},
  \bibinfo{journal}{SSRN eLibrary, Inertia of Social Rotation in Laboratory 2x2
  Population Games, http://dx.doi.org/10.2139/ssrn.1985513}
  (\bibinfo{year}{2012}{\natexlab{a}}).

\bibitem[{\citenamefont{Oprea et~al.}(2011)\citenamefont{Oprea, Henwood, and
  Friedman}}]{Friedman2011}
\bibinfo{author}{\bibfnamefont{R.}~\bibnamefont{Oprea}},
  \bibinfo{author}{\bibfnamefont{K.}~\bibnamefont{Henwood}}, \bibnamefont{and}
  \bibinfo{author}{\bibfnamefont{D.}~\bibnamefont{Friedman}},
  \bibinfo{journal}{Journal of Economic Theory} \textbf{\bibinfo{volume}{146}},
  \bibinfo{pages}{2206} (\bibinfo{year}{2011}).

\bibitem[{\citenamefont{Xu and Wang}(2011{\natexlab{b}})}]{XuWang2011ICCS}
\bibinfo{author}{\bibfnamefont{B.}~\bibnamefont{Xu}} \bibnamefont{and}
  \bibinfo{author}{\bibfnamefont{Z.}~\bibnamefont{Wang}},
  \emph{\bibinfo{title}{Evolutionary Dynamical Pattern of \"Coyness and
  Philandering\": Evidence from Experimental Economics}}, vol.
  \bibinfo{volume}{VIII} (\bibinfo{publisher}{NECSI Knowledge Press, ISBN
  978-0-9656328-4-3}, \bibinfo{year}{2011}{\natexlab{b}}).

\bibitem[{\citenamefont{Hoffman et~al.}(2012)\citenamefont{Hoffman, Suetens,
  Nowak, and Gneezy}}]{nowak2012}
\bibinfo{author}{\bibfnamefont{M.}~\bibnamefont{Hoffman}},
  \bibinfo{author}{\bibfnamefont{S.}~\bibnamefont{Suetens}},
  \bibinfo{author}{\bibfnamefont{M.}~\bibnamefont{Nowak}}, \bibnamefont{and}
  \bibinfo{author}{\bibfnamefont{U.}~\bibnamefont{Gneezy}},
  \bibinfo{journal}{An experimental test of Nash equilibrium versus
  evolutionary stability}  (\bibinfo{year}{2012}).

\bibitem[{\citenamefont{Xu and Wang}(2012{\natexlab{b}})}]{XuWang2012LDF}
\bibinfo{author}{\bibfnamefont{B.}~\bibnamefont{Xu}} \bibnamefont{and}
  \bibinfo{author}{\bibfnamefont{Z.}~\bibnamefont{Wang}},
  \bibinfo{journal}{SSRN eLibrary, Symmetry Properties of the Large-Deviation
  Function of the Social Rotation in Laboratory Population Games,
  http://dx.doi.org/10.2139/ssrn.1988985}
  (\bibinfo{year}{2012}{\natexlab{b}}).

\bibitem[{\citenamefont{Xu and Wang}(2011{\natexlab{c}})}]{XWMm2011}
\bibinfo{author}{\bibfnamefont{B.}~\bibnamefont{Xu}} \bibnamefont{and}
  \bibinfo{author}{\bibfnamefont{Z.}~\bibnamefont{Wang}},
  \bibinfo{journal}{SSRN eLibrary, Social Transition Spectrum in Constant Sum
  2x2 Games with Human Subjects, http://dx.doi.org/10.2139/ssrn.1910045}
  (\bibinfo{year}{2011}{\natexlab{c}}).

\bibitem[{\citenamefont{Xu et~al.}(2012)\citenamefont{Xu, Zhang, Wang, and
  Zhang}}]{xuetal2012Maxent}
\bibinfo{author}{\bibfnamefont{B.}~\bibnamefont{Xu}},
  \bibinfo{author}{\bibfnamefont{H.}~\bibnamefont{Zhang}},
  \bibinfo{author}{\bibfnamefont{Z.}~\bibnamefont{Wang}}, \bibnamefont{and}
  \bibinfo{author}{\bibfnamefont{J.}~\bibnamefont{Zhang}},
  \bibinfo{journal}{Physics Letters A,
  http://dx.doi.org/10.1016/j.physleta.2012.02.047}  (\bibinfo{year}{2012}).

\bibitem[{\citenamefont{Cason et~al.}(2012)\citenamefont{Cason, Friedman, and
  Hopkins}}]{Friedman2012}
\bibinfo{author}{\bibfnamefont{T.}~\bibnamefont{Cason}},
  \bibinfo{author}{\bibfnamefont{D.}~\bibnamefont{Friedman}}, \bibnamefont{and}
  \bibinfo{author}{\bibfnamefont{E.}~\bibnamefont{Hopkins}},
  \bibinfo{journal}{Cycles and Instability in a Rock-Paper-Scissors Population
  Game: a Continuous Time Experiment}  (\bibinfo{year}{2012}).

\bibitem[{\citenamefont{Tang}(2010)}]{Tang2010}
\bibinfo{author}{\bibfnamefont{F.}~\bibnamefont{Tang}}, \bibinfo{journal}{The
  Selten School of Behavioral Economics} pp. \bibinfo{pages}{33--49}
  (\bibinfo{year}{2010}).

\bibitem[{\citenamefont{Xu and Wang}(2011{\natexlab{d}})}]{XuWang2011}
\bibinfo{author}{\bibfnamefont{B.}~\bibnamefont{Xu}} \bibnamefont{and}
  \bibinfo{author}{\bibfnamefont{Z.}~\bibnamefont{Wang}},
  \bibinfo{journal}{Arxiv preprint arXiv:1105.3433}
  (\bibinfo{year}{2011}{\natexlab{d}}).

\end{thebibliography}

\end{document}